\documentclass[twocolumn]{aastex631}
\usepackage{gensymb}
\usepackage{CJKutf8}
\newcommand{\kms}{\ifmmode{\,\hbox{km\,s}^{-1}}\else {\rm\,km\,s$^{-1}$}\fi}

\shorttitle{C-19 and Hot, Wide, Streams}
\shortauthors{Carlberg, et al.}

\begin{document}

\title{C-19 and Hot, Wide, Star Streams}

\author[0000-0002-7667-0081]{Raymond G. Carlberg}
\affiliation{Department of Astronomy \& Astrophysics,
University of Toronto,
Toronto, ON M5S 3H4, Canada} 
\email{raymond.carlberg@utoronto.ca}

\author[0000-0002-3292-9709]{Rodrigo Ibata}
\affiliation{Université de Strasbourg, CNRS, Observatoire astronomique de Strasbourg, UMR 7550, F-67000 Strasbourg, France}

\author[0000-0002-1349-202X]{Nicolas F. Martin}
\affiliation{Université de Strasbourg, CNRS, Observatoire astronomique de Strasbourg, UMR 7550, F-67000 Strasbourg, France}

\author[0000-0003-4501-103X]{Else Starkenburg}
\affiliation{Kapteyn Astronomical Institute, University of Groningen, Landleven 12, 9747 AD Groningen, The Netherlands}

\author[0000-0001-5200-3973]{David S. Aguado}
\affiliation{Instituto de Astrof\'{\i}sica de Canarias, V\'{\i}a L\'actea, 38205 La Laguna, Tenerife, Spain}

\author[0000-0002-8318-433X]{Khyati Malhan} 
\affiliation{DARK, Niels Bohr Institute, University of Copenhagen, Jagtvej 128, 2200 Copenhagen, Denmark}

\author[0000-0003-4134-2042]{Kim Venn}
\affiliation{Department of Physics and Astronomy, University of Victoria, Victoria, BC, V8W 3P2, Canada}

\author[0000-0002-8129-5415]{Zhen Yuan (\begin{CJK}{UTF8}{gbsn}   袁珍 )\end{CJK}}
\affiliation{School of Astronomy and Space Science, Nanjing University,
Key Laboratory of Modern Astronomy and Astrophysics (Nanjing University), Ministry of Education, Nanjing 210093, China}

\begin{abstract}
The C-19 star stream has the abundance characteristics of an unusually metal poor globular cluster but kinematically is uncharacteristically hot and wide for a cluster stream, having a line of sight velocity dispersion of $7\pm 2$ \kms\  and a 1-sigma width of  240 pc. We show that the tidal dissolution of an old, lower mass,  globular cluster in a CDM galactic halo can create  a hot, wide stream currently near orbital apocenter.   A cosmological Milky Way n-body simulation motivates the parameters for an evolving Milky Way halo potential containing an orbiting subhalo population in which we model a star cluster progenitor of C-19. The same model parameters have been used for a GD-1 stream model.  The $\sim 7 \kms$ velocity dispersion is readily accomplished with an evolving CDM subhalo population, a progenitor cluster mass $\simeq 2\times 10^4 M_\odot$ and an orbit that keeps the progenitor orbital pericenter within about 10 kpc of the Milky Way dark halo or its precursors.
\end{abstract}

\section{INTRODUCTION}

The C-19 stream of extremely metal poor stars  \citep{C19Nature} was discovered \citep{Ibata21} in the Gaia data \citep{GaiaMission} using the STREAMFINDER algorithm \citep{Malhan18}. High resolution spectroscopy \citep{C19Nature,Yuan22} of photometrically identified low metallicity stream candidates \citep{pristine} established that C-19 has a metal abundance of [Fe/H] = -3.4 with a 95\% confidence [Fe/H] spread of  0.18 \citep{C19Nature}. Further investigation has found additional stream candidates giving a length of more than 100 degrees \citep{Yuan22,Viswanathan24,Yuan25}. The  spread in stellar abundances is  close to the detection limit and the abundance patterns are characteristic of globular cluster stars \citep{Yuan22}. The  abundances and extrapolated luminosity of approximately $10^4 L_\odot$ led \citet{C19Nature} to argue that the progenitor was more likely to be a disrupted globular cluster than a disrupted dwarf galaxy,  while considering that the 1-sigma width of 240 pc, and line of sight velocity dispersion of 6 \kms\ were more in keeping with a disrupted dwarf.  A dynamical model of a disrupting dark matter dominated  dwarf  galaxy can account for the kinematic properties of the stream \citep{Errani22}.

The question addressed here is under what conditions can a globular cluster progenitor be compatible with the C-19 kinematic data. Most of the known streams in the compilation \href{https://github.com/cmateu/galstreams}{Galstreams}  \citep{Mateu23} are  thinner than C-19 with typical FWHM of 100 pc or less. On the other hand, simulations of globular cluster streams in realistic cosmological conditions find that the dark matter subhalos present in the galactic halo cause the velocity dispersion of stream stars to increase with time \citep{CA23,Carlberg24}.  Both the width and velocity distribution of streams have a core with extended wings, aka cocoon \citep{Carlberg18,Malhan19}.  

A low velocity dispersion tidal stream from a globular cluster is sensitive to small scale potential perturbations along its orbit. The velocity at which stars join a stream determines the width of the low velocity dispersion core. Subsequent encounters with orbiting dark subhalos perturb stars into the extended wings of the velocity and width distribution. Individual stream stars have a range of orbital speeds which, with time, spreads the perturbations along the stream. The growing dataset for the GD-1 stream provides evidence for the existence of a core-wing velocity and width structure \citep{Ibata24,Valluri24,Carlberg25}.   The C-19 stream is on a nearly polar, inner halo orbit, 8-24 kpc, similar to GD-1,  so should also show significant subhalo heating.   

In \S\ref{data_orbit} the  velocity dispersion of various C-19 subsamples along the stream is calculated relative to a quadratic fit. The stream data is also used to find a progenitor star cluster orbit in the model potential.   Two types of stream simulation are discussed. First, a $\Lambda$CDM cosmological simulation with dark matter particles and star particle clusters provides an overview of the kinematic properties of streams from low mass star clusters. In \S\ref{sec_sims}  the C-19 velocity dispersion and width are compared to those of the streams from a population of dissolving star clusters in a CDM cosmological simulation. The cosmological simulation provides the primary halo mass and subhalo masses and orbits with time for a second type of simulation, using an evolving potential with subhalos. A low mass n-body star cluster on the orbit of C-19 is followed in this potential \S\ref{c19_model}, which is the same as the model used for GD-1 \citep{Carlberg25}. The results are  discussed \S\ref{sec_discussion} to assess the accuracy of the model and its ability to constrain the subhalo content of the Milky Way. Other currently known wide streams from globular clusters are briefly discussed along with the likelihood that there are other wide streams yet to be discovered.

\section{C-19 Data and Orbit\label{data_orbit}}

\begin{figure}
\begin{center}
\includegraphics[angle=0,scale=0.42,trim=60 0 0 35, clip=true]{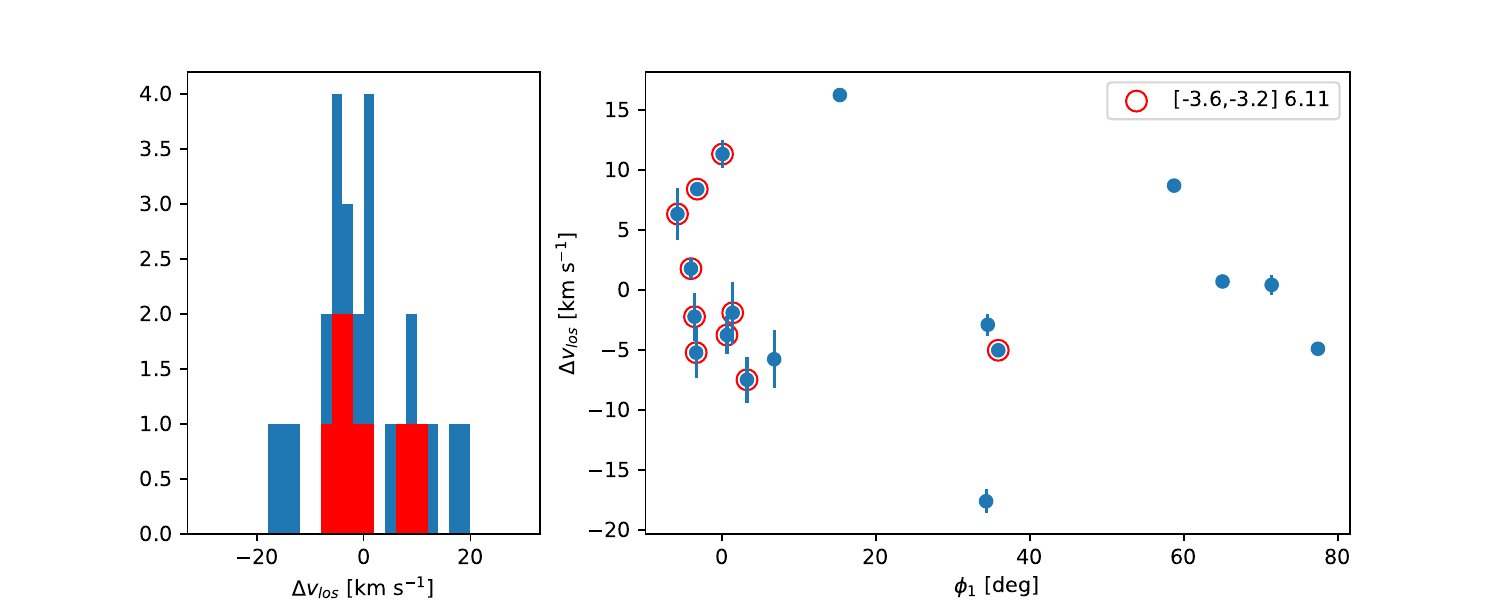}
\end{center}
\caption{The C-19 velocity offsets from a best fit quadratic (left panel) and the velocity offsets along the stream (right panel). The highest confidence member stars, having [Fe/H] in the [-3.6,-3.2] range, are marked in red. 
}
\label{fig_data}
\end{figure}

The sparse velocity measurements along the 100\degree\ length of C-19 limit the accuracy of C-19 velocity dispersion measurements.  \citet{C19Nature,Yuan22} found $6.2^{+2.0}_{-1.4} \kms$ and \citet{Yuan25} $11.1^{+1.9}_{-1.6} \kms$ relative to an orbit.   We use the minimalist approach of simply fitting a quadratic to the measured line of sight velocities and then calculating the velocity offsets from the quadratic as shown in Figure~\ref{fig_data}. The resulting velocity dispersion values are given in Table~\ref{tab_c19sig}. We will use the values in the [Fe/H] =[-3.6, -3.1] range for $\phi_1=[-5,5]$ and for all $\phi_1$, which are $6.2\pm 2.3$ and $7.6\pm 1.8$ \kms, respectively.

We use the data of \citet{Yuan25} to define a C-19 orbit. C-19 stream coordinates \citep{Ibata24} have a pole of RA=81.45\degree\ Dec= -6.346\degree\ and a  RA zero of 354.356\degree\ \citep{Ibata24}. The line of sight velocities are fit to a quadratic, with the value at $\phi_1=0$ used as the starting point, along with the means of the values within $|\phi_1| < 10\degree$ resulting in a starting position in ICRS coordinates of (ra, dec, distance) in (deg, deg, kpc)     (354.691, 24.468, 18.75) and  (pmra-cosdec, pmdec, radial-velocity) in (mas yr$^{-1}$, mas yr$^{-1}$, \kms)     (1.22, -2.94, -190.2).  

\begin{figure}
\begin{center}
\includegraphics[angle=0,scale=0.52,trim=25 30 0 60, clip=true]{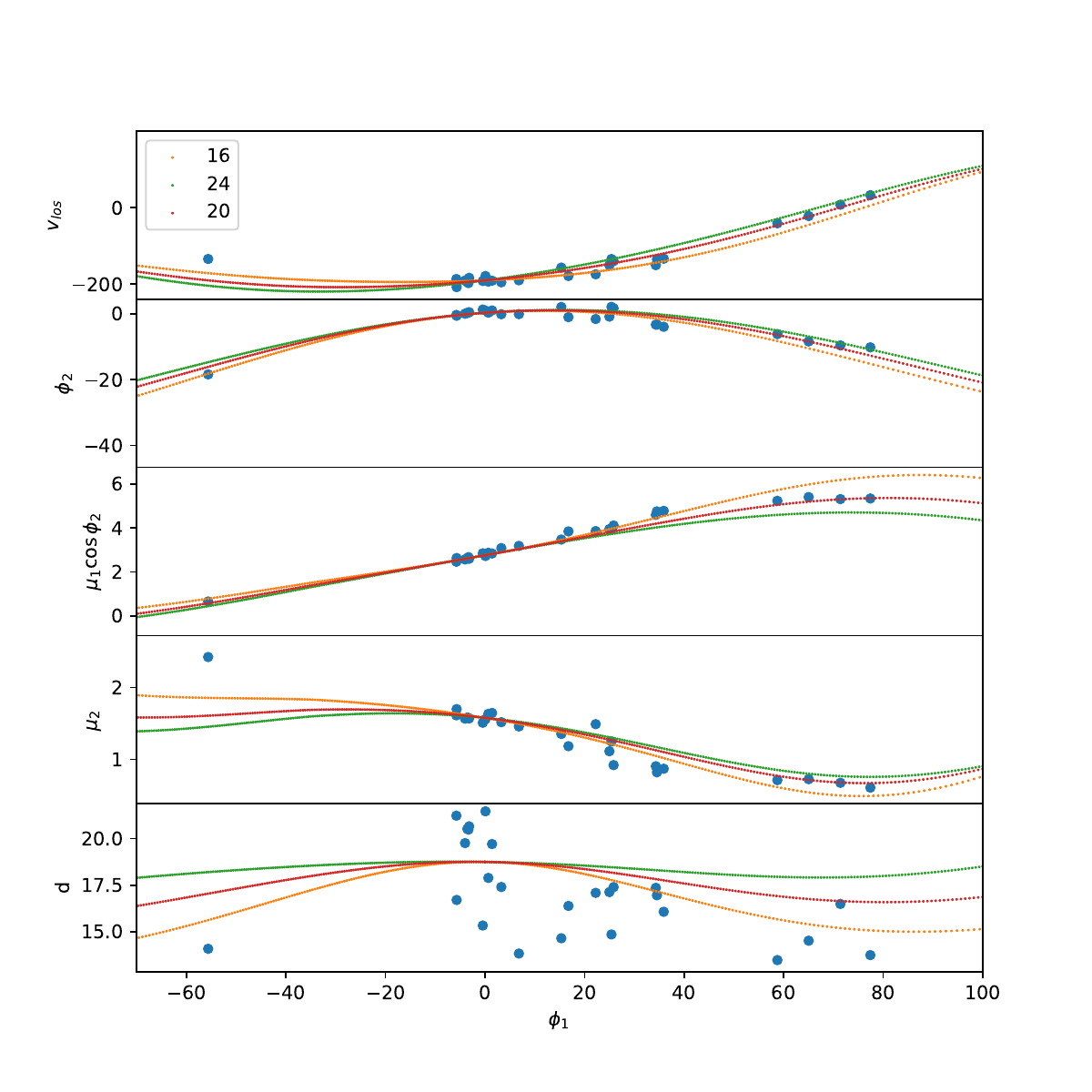}
\end{center}
\caption{Orbits for the progenitor star cluster varying the scale radius of the halo potential from 16 to 24 kpc. 
}
\label{fig_modorbits}
\end{figure}

The orbit segment presented in Figure~\ref{fig_modorbits} uses a potential composed of a Miyamoto-Nagai disk \citep{MN75} and an oblate, mildly triaxial NFW halo \citep{NFW} taken from a Milky Way-like cosmological simulation \citep{Carlberg24} that closely approximates the mass profile measurements of \citet{Shen22}, with a mass $9.22 \times 10^{11} M_\odot$ and concentration of 13.84.  The orbits for three values of the NFW scale radius are shown in Figure~\ref{fig_modorbits}.  A 20 kpc scale radius provides the best approximation to the data although the measurement at $\phi_1 = -55.6\degree$ is significantly offset. The resulting orbit has a pericenter of 10.0 kpc, apocenter of 22.9 kpc and current galactocentric distance of 21.7 kpc. That is, the cluster of points near $\phi_1=0$ are just past the orbit apocenter. The orbital apocenter is near the stream $\phi_1$=-20\degree\ and the pericenter is near $\phi_1$=110\degree. A stream from a cluster on this orbit will have stars systematically offset from the cluster orbit.

\begin{table} 
\caption{C-19 Line of Sight Velocity Dispersion
\label{tab_c19sig}}
 \begin{tabular}{|r|r|r|r|} 
 \hline
 $\phi_1$ & [Fe/H] & N & $\sigma_{los}$ ~~~\\ 
 \hline
 degree   &      ~~     &  ~~    & \kms\ \\
 \hline
  all & $>$ -4 & 26 & $9.2 \pm 1.8$ \\
  all & [-3.6,-3.1] & 19 & $7.6 \pm 1.8 $ \\
 -5, 5 &  [-3.6,-3.1] & 8 & $ 6.2 \pm 2.3 $ \\
 -5, 5 & all & 9 & $5.9 \pm 2.1$ \\
 $>$ 30 &  [-3.6,-3.1] & 7 & $7.4 \pm 3.0$ \\
\hline
\end{tabular}
\end{table}

\section{Low Mass Star Clusters in Cosmological Simulations\label{sec_sims}}

\begin{figure}
\begin{center}
\includegraphics[angle=0,scale=0.7,trim=10 10 0 0, clip=true]{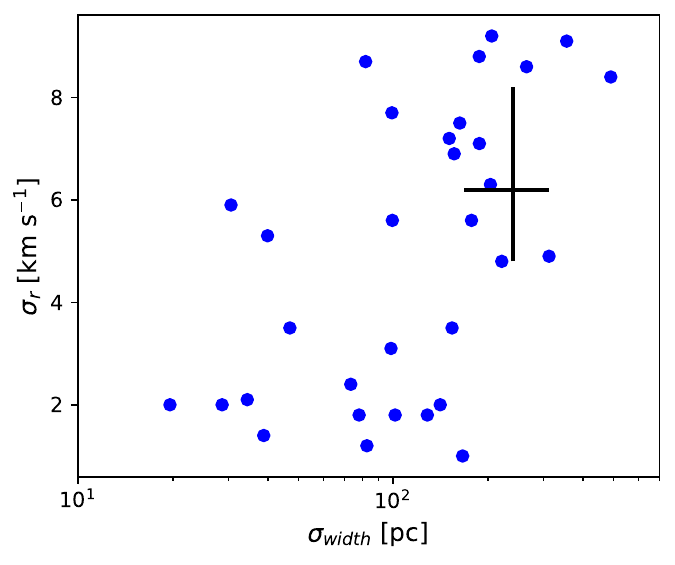}
\end{center}
\caption{The stream radial velocity dispersion (galactocentric) vs stream width as measured perpendicular to the stream track for low mass cluster progenitors in CDM cosmological simulations at the final time.  The error bar gives the C-19 values from \citet{Yuan22}.
}
\label{fig_sigwidth}
\end{figure}

The low mass cluster streams that develop in a cosmological Milky Way-like halo seeded with progenitor star clusters in the mass range 5-20$\times 10^3 M_\odot$ provide useful context for the C-19 stream and provide the parameters for a evolving potential model with no dark matter particles. The simulation is identical to those in \citet{Carlberg24}  other than having smaller initial star cluster masses and was used for GD-1 \citep{Carlberg25}.  In brief, a low resolution simulation in 50/h Mpc box set up with $\Lambda$CDM initial conditions is run to redshift zero and a Milky Way like halo is identified.  A 2 Mpc (co-moving) region that evolves into the Milky Way like halo is reinitialized at redshift 50  with 10322 $M_\odot$ dark matter particles \citep{MUSIC}.  The region is run to 1 Gyr with Gadget4 \citep{Gadget4} at which time the dark matter subhalos are identified with the AHF halo finder \citep{AHF1,AHF2}. The subhalos more massive than $3\times 10^8 M_\odot$ have n-body star clusters with a half mass radius of $\sim 5$ pc composed of 1 $M_\odot$ star particles added into the selected subhalos to mimic a high redshift dwarf galaxy population of globular clusters. The dark matter particles with the added star particle clusters is then evolved to redshift zero, with a Miyamoto-Nagai \citep{MN75} disk started to at 5 Gyr linearly increasing in mass in the center. The internal dynamics of the star clusters causes them to lose mass \citep{BT08}, which the tidal fields pull away into streams. The width and velocity dispersion of the streams are measured in galactocentric great circle coordinates. We select the streams that have  a minimum angular momentum of 2000 kpc-\kms,  which excludes  those that interact  with the disk, and streams orbiting  within 60 kpc, to be comparable to currently observable streams. Figure~\ref{fig_sigwidth} shows the stream radial velocity dispersion (galactocentric) and stream width for the streams in the CDM simulation along with the C-19 values as reported in \citet{Yuan22}.  About a third of the streams have velocity dispersion values of C-19 or larger. 

\begin{figure*}
\begin{center}
\includegraphics[angle=0,scale=0.65,trim=0 10 0 0, clip=true]{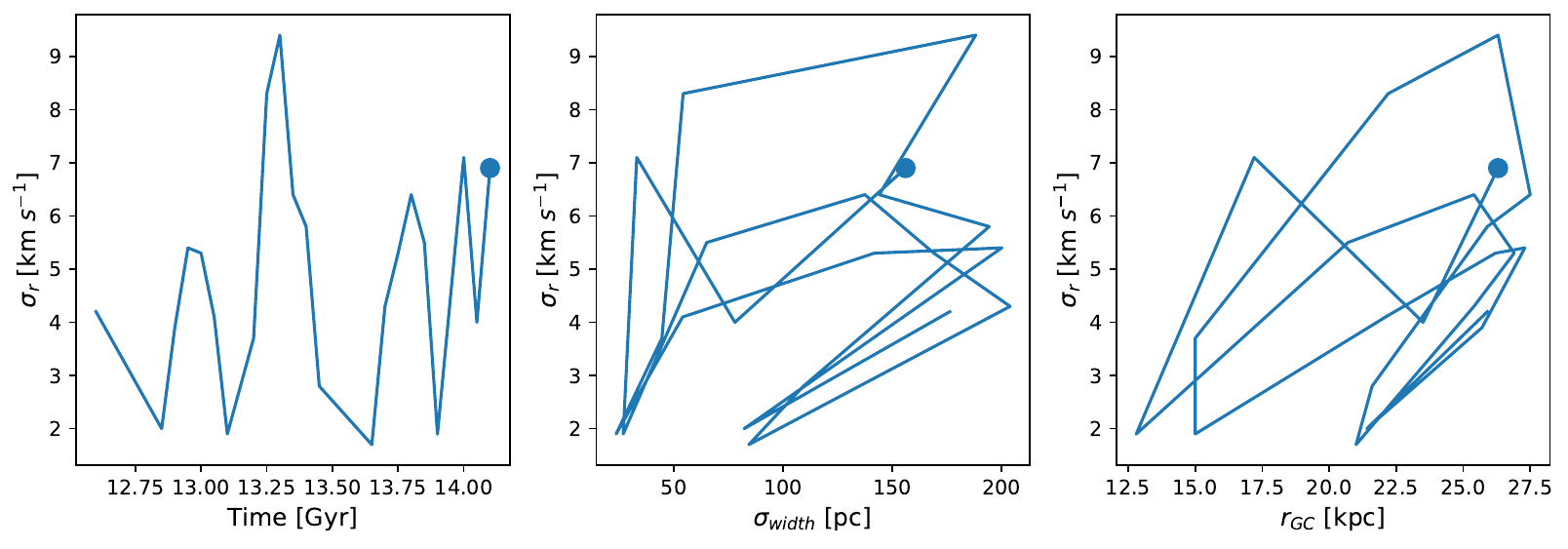}
\end{center}
\caption{The orbital evolution of the stream radial velocity dispersion (left panel) with time for a CDM stream selected to have a high final velocity dispersion and an orbit comparable to C-19. The final time is marked with a dot. The middle panel shows the evolution of velocity dispersion with width. The left panel shows the evolution of the radial velocity dispersion with galactocentric distance.
}
\label{fig_orbitsw}
\end{figure*}

Figure~\ref{fig_orbitsw} shows the time dependence of the radial velocity dispersion and stream width for a stream in the  cosmological simulation on an orbit comparable to C-19 (cluster number 154). The left panel shows the time variation of the radial velocity dispersion, $\sigma_r$, in the stream. The right hand panel shows the relation between  $\sigma_r$ and the galactic radius of the center of  the streams. The middle panel shows the stream width-$\sigma_r$ relation. The stream width and $\sigma_r$ increase as the stream orbits through apocenter, narrowing again as it moves back to a pericenter. A movie of this stream is at \href{https://www.astro.utoronto.ca/~carlberg/streams/417}{CDM Low Mass Clusters}.  In a simple non-evolving potential the distribution of stream stars projected onto the orbital plane varies from a narrow distribution at pericenter to extended ``feathers'' at apocenter, as illustrated in Figure~2 of \citet{Carlberg15a}. The orbital variation is a consequence of the galactic tidal field accelerating stars away from the cluster near pericenter, the subsequent orbital  spreading of the unbound stars at apocenter, then regrouping at pericenter again, reminiscent of the orbital dependence of velocities of an accretion remnant \citep{HW99}.

\section{C-19 Stream Models \label{c19_model}}

A detailed model prediction for C-19, or any other specific stream, needs to reproduce the current orbital kinematics and the width and velocity spread of the stream. It should also reproduce the statistical characteristics of the density variations along the stream, which we will address elsewhere. The usual approach is to specify a Milky Way potential, backward integrate the orbit to a desired starting time, place a stream producing star cluster at the starting point and then integrate it forward in the model potential. Integrating such a model even for 10 Gyr in a smooth model potential fails to produce a stream with the width and velocity dispersion of C-19 \citep{Errani22}. Since the stream heating in the cosmological simulation is primarily due to the subhalos \citep{Carlberg24}, adding the subhalos as found in the cosmological simulation to a model potential should be able to  heat the C-19 stream to the measured value. 

\subsection{Evolving Potential Model}

Our C-19 stream model uses a precomputed time varying galactic potential, the same as used for GD-1 \citep{Carlberg25}. The model allows us to accurately calculate the development of any desired stream, has no star particle heating from dark matter particles, captures the mass buildup of the primary halos but not the asymmetric potential fluctuations as mergers occur. Inserted into the evolving potential model is a n-body star cluster with  mass of $1-5\times 10^4 M_\odot$, composed of $1 M_\odot$ star particles.  The clusters begin as King models \citep{King66} with half-mass radii  in the range 3-8 pc for the results reported here. The star particles have a softening of 1 pc and are integrated with the Gadget4 code \citep{Gadget4} using the same method as in \citet{Carlberg24}. Two body star-star interactions that are central to the evolution of globular clusters are included by adding small random velocities calculated from the relaxation time \citep{Spitzer87,BT08} to  star particles within the virial radius, about 5 pc,  every 5 Myr \citep{CA23}. The tidal field of the galactic potential pulls stars away from the outskirts of the star cluster and into the star stream with no model assumptions. 

The star cluster orbits in potential modeled after the evolution of the growing Milky Way like halo of the cosmological  simulation. The same potential as in \citet{Carlberg25} for the GD-1 stream is used here, with only the star cluster orbit changed. The primary mass component  is an evolving, triaxial NFW potential, with mass, $9.2\times 10^{11} M_\odot$, virial radius 200 kpc, concentration 13.84 and density triaxiality from the AHF halo finder \citep{AHF1,AHF2} measurements at the final moment the simulation. The potential triaxiality is set to b/a=0.97, c/a=0.83. A Miyamoto-Nagai galactic bulge-disk \citep{MN75}  grows within the halo to a final mass $6.8\times 10^{10} M_\odot$, with a = 3.0 kpc and b= 0.28, the same model parameters adopted in the cosmological simulations of \citet{Carlberg24}.  The final time potential is identical to the one used for the orbit segment in \S\ref{data_orbit}. The framework could have a more complex buildup of the primary halo and add a rotating galactic bar if required.  

The positions and velocities of the population of subhalos above $10^6 M_\odot$ as found in the cosmological simulation at the chosen starting time is integrated in the combined Milky Way dark halo and disk. The subhalos do not interact with each other. The subhalos' masses decrease with an exponential decay on a timescale of 5.5 Gyr starting at 5 Gyr, which leads to a factor of five decrease in overall masses to redshift zero. The decrease in subhalo mass with time means that the subhalo $N(>M)\propto M^{-1}$ relation for CDM subhalos \citep{Springel08} within the main halo declines approximately as measured in the fully dynamical cosmological simulations.  Subhalos at the start time above a minimum mass of $10^6 M_\odot$ to a maximum $3\times 10^8 M_\odot$  are included in the evolving potential model. Lower mass halos have essentially no dynamical effect on the stream and subhalos above  $3\times 10^8 M_\odot$ are expected to contain visible dwarf galaxies, both of which interact with streams so rarely that they are not important for stream structure development. This universal subhalo evolution  is somewhat conservative,  since it does not include the halos above $3\times 10^8 M_\odot$ at the start or $6\times 10^7 M_\odot$ at the final time. As a test of model sensitivity we ran one case with the start time mass limit boosted to $15\times 10^8 M_\odot$ which becomes $3\times 10^8 M_\odot$ at the end which adds 113 subhalos to the original 8004, with the most massive subhalos usually at radii beyond the orbit of C-19.

The subhalos are Hernquist spheres with a scale radius set to  the radius of the maximum of the circular velocity, $r_{max}$, that AHF finds. As the subhalos lose mass their scale radii are adjusted as $[M/M(0)]^{1/3}$ to keep the characteristic density of each subhalo constant. In addition to the subhalos the 55 dwarf galaxies with kinematics \citep{McConnachie12} are included. The dwarf galaxy dark halos are assigned maximal masses, $M= 5\times 10^8 (L/L_\odot)^{0.65} M_\odot$, which roughly matches the high mass end of the subhalo mass distribution function. The dwarf galaxy halo  radii are set to $a = 1.0 (M/10^8 M_\odot)^{0.23}$ kpc, using a fit to the $r_{max}$ mass relation of the subhalos in the simulations.   All the subhalo orbits are precomputed in the evolving primary potential to generate spline coefficients which the n-body star cluster uses as an external potential.

\subsection{C-19 Stream Models}

The progenitor center is placed  along the stream track at a stream angle chosen to be between $\phi_1$ of -40\degree\ and 0\degree, then integrated backward in the evolving potential with the evolving subhalo distribution to the desired start time. The model star cluster is then placed at that position and integrated  forward with Gadget4.  The final time progenitor location depends on the details of encounters with subhalos, so it varies from run to run, generally coming back to slightly positive $\phi_1$ values. Decreasing the time error tolerance a factor of two leads to essentially identical orbital outcomes.

The total stellar mass of C-19 is estimated to be  in a wide range  of $3\times 10^3$ to $5 \times 10^4 M_\odot$ \citep{C19Nature,Yuan22,Ibata24,Yuan25}. There is no visible progenitor, as expected for an old star cluster with a half mass relaxation time of 1-2 Gyr orbiting in the Milky Way \citep{Gnedin97,BT08,Errani22}. The mass loss  timescale of the star cluster increases with mass, for a fixed density, so plays a role in determining the structure of the stream. The simulations presented here start with star cluster masses of $1, 2$ and $5\times 10^4 M_\odot$. More massive progenitor star clusters of the same density take longer to dissolve, as do lower density clusters, which then have less time to heat, leading to lower velocity dispersion streams, all else being equal.

\begin{figure}
\begin{center}
\includegraphics[angle=0,scale=0.5,trim= 65 180 30 160, clip=true]{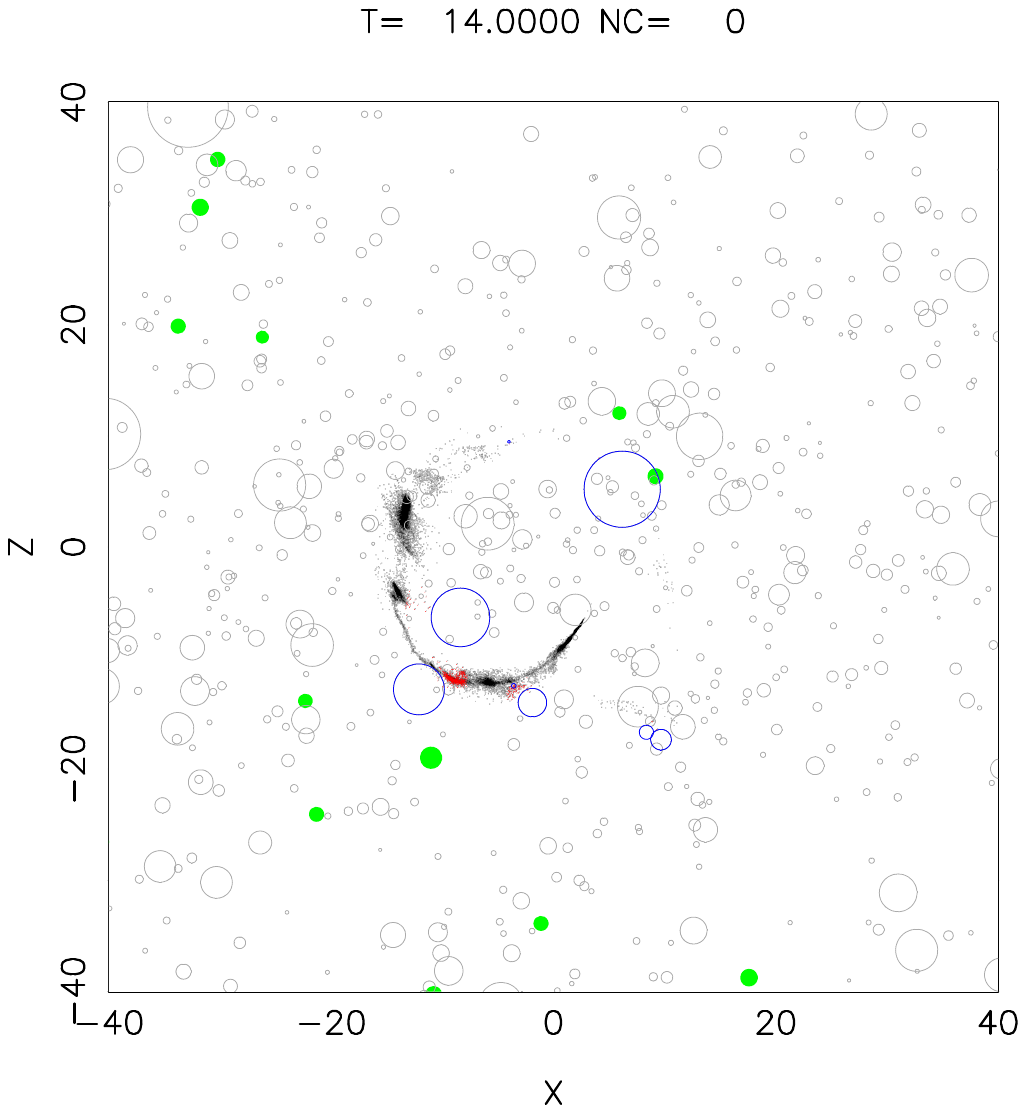}
\includegraphics[angle=0,scale=0.475,trim=55 150 30 160, clip=true]{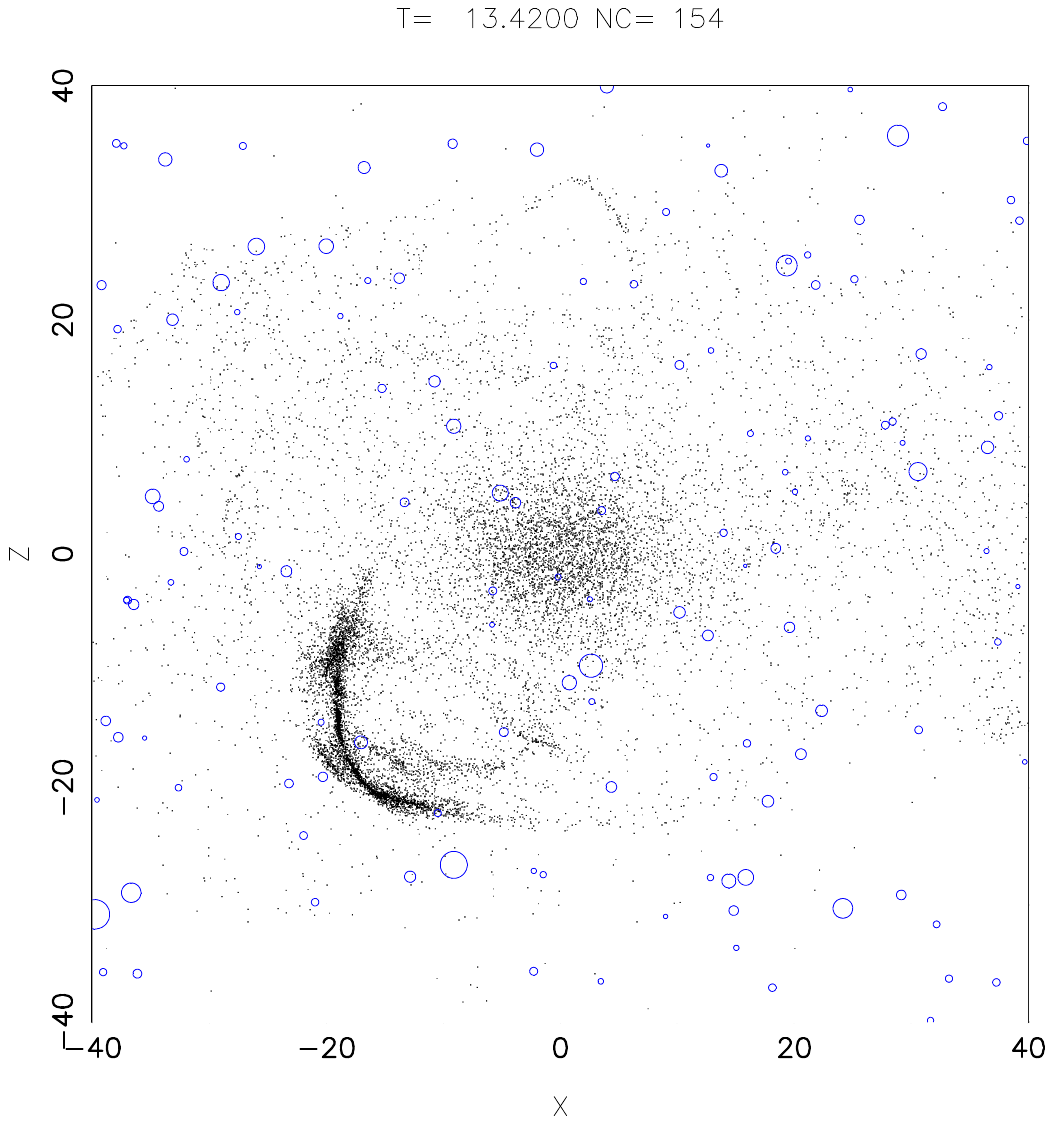}
\end{center}
\caption{(top) The XZ projection in kpc of a CDM evolving potential model simulation of C-19 at the final time.  The stream is shown in greyscale, the dwarf galaxies are filled green circles and the subhalos are grey circles.
(bottom) The XZ projection of a stream on an orbit similar to C-19 in the cosmological CDM simulation at time 13.95, which has a stream orbital position similar to the simulation of  the top panel, rotating in the opposite direction. Movies of the evolving potential C-19 simulations and the cosmological simulation with streams are at \href{https://www.astro.utoronto.ca/~carlberg/streams/c19}{C-19 movies} and  \href{https://www.astro.utoronto.ca/~carlberg/streams/417}{CDM Low Mass Clusters}, respectively.
}
\label{fig_simz}
\end{figure}

Figure~\ref{fig_simz} top panel shows  an XZ projection of a snapshot of the star particles along with the CDM subhalos at a cosmological time of 14.0 Gyr, 0.1 Gyr before the present time to show a time when the stream is interacting with some subhalos. For a CDM cosmology  8004 subhalos are followed initially within 200 kpc. Most of the subhalos orbit well outside the region where C-19 orbits but they are included for completeness. The dwarf galaxies are shown in green in Figure~\ref{fig_simz} although interactions are rare and usually do not influence inner halo streams  \citep{Bonaca19}. The XY projection of the stream is shown in grayscale at time 1 Gyr before present when stream particles are still passing through orbital apocenter. Subhalos that are within 2 scale radii of the stream are shown in blue and the stream particles within that radius are red. The interacting subhalo near the center of the image has a current mass of $2.2\times 10^7 M_\odot$ and induces a velocity change of about 0.5 \kms\ in the nearest stream particles. The interactions that dominate the stream velocity variations are largely from subhalos halos in the mass decade around $10^{7.5} M_\odot$ at times between 5 and 10 Gyr of the simulation, once the stream tidal tails have acquired appreciable length but before the subhalo interactions diminish as their numbers at a given mass decline. 

To help validate the evolving potential model as representative of the behavior of a stream in the cosmological n-body model, a stream on an orbit similar to C-19 along with the subhalos from the cosmological simulation is shown in the lower panel of Figure~\ref{fig_simz}. An approximate match to the phase of the C-19 orbit is shown to demonstrate the width, length and structural similarity. The subhalo populations are identical at the start time. Although they differ at the displayed time, subhalo heating is not effective at late times. The darker, higher density parts of the streams have broadly similar features, with density variations along their length and star particles spread around a narrower core. The significant difference is that  fully cosmological stream of Figure~\ref{fig_simz} has a low density population of star particles scattered over the entire inner halo, a consequence of the large scale potential fluctuations that occur during accretion \citep{Carlberg24} separating stream segments onto new orbits as shown in the movies available at \href{https://www.astro.utoronto.ca/~carlberg/streams/417}{CDM Low Mass Clusters}. 

\begin{figure}
\begin{center}
\includegraphics[angle=0,scale=0.33,trim=60 0 30 20, clip=true]{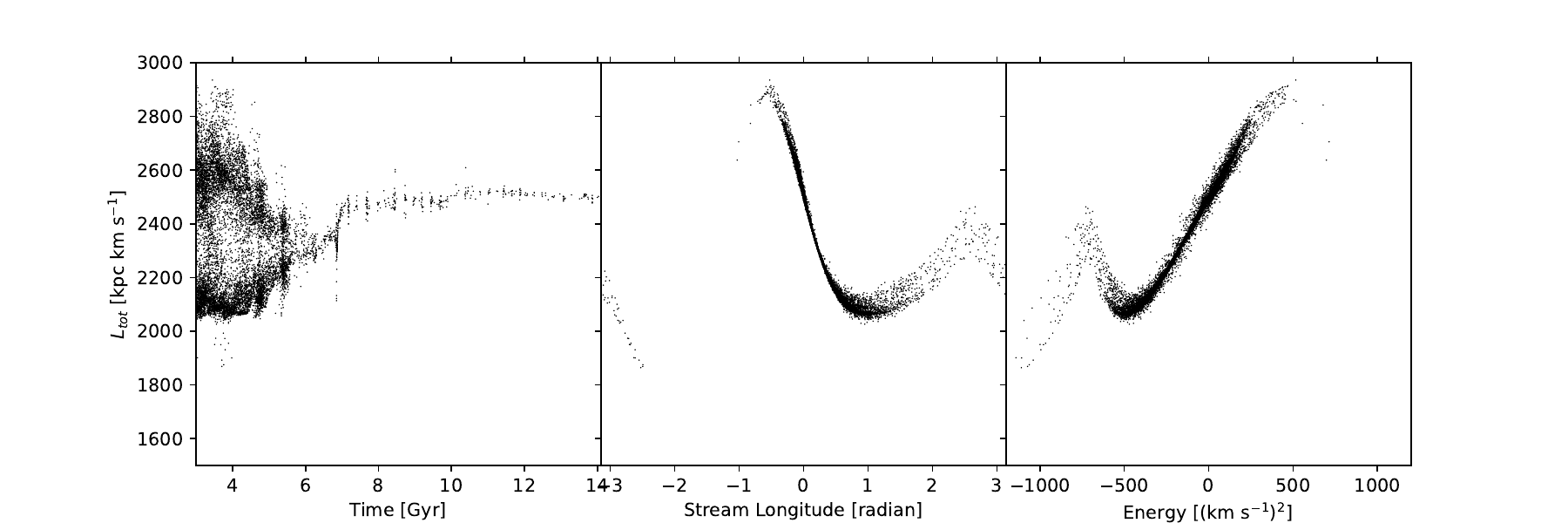}
\includegraphics[angle=0,scale=0.33,trim=60 0 30 20, clip=true]{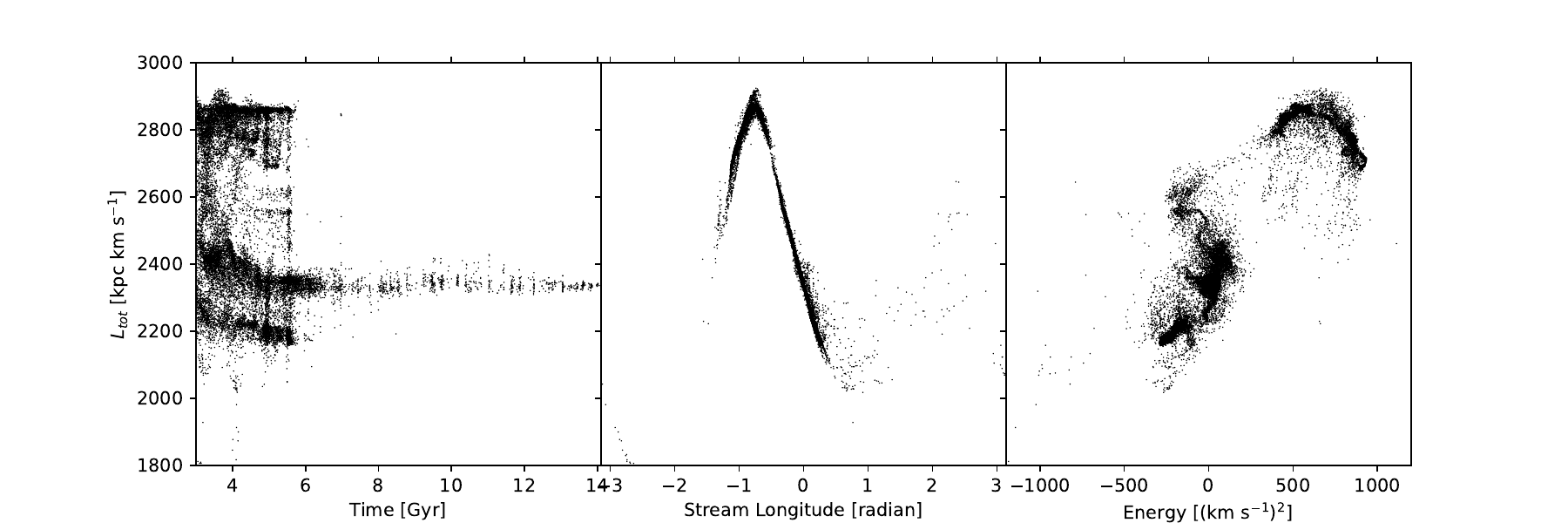}
\end{center}
\caption{The top row of panels is for a simulation with no subhalos, the bottom for a CDM subhalo simulation. The star clusters start with a mass of $2 \times 10^4 M_\odot$  $r_h(0)=3.3$ pc.  In each row the left panels show the final total angular momentum, L, against the time the star particle left the cluster, the middle panels show L with stream galactocentric great circle longitude relative to the nominal dissolved cluster location, and the right panels show L with the orbital energy relative to the energy of the cluster center position. 
}
\label{fig_LE}
\end{figure}

The internal kinematics of a stream blurs out lower mass subhalo interactions with a stream. The left panel of Figure~\ref{fig_LE} shows the total angular momentum (not a conserved quantity in this potential) of the star cluster particles measured at the final time  against the time at which the star particles left the cluster to join the stream, defined as the time when the  particles are more than 20 pc from the cluster center for the first time. The spread between the trailing (high angular momentum) and leading (low angular momentum) branches of the tidal stream decreases with time as the cluster loses mass causing the tidal radius to move closer to the cluster. The simulation with no subhalos retains the simple pericenter mass loss pattern to the end of the simulation, whereas the CDM subhalo model becomes increasingly perturbed. The total angular momentum (not a conserved quantity in the model potential) with stream longitude in the middle panel shows some increasing substructure with increasing numbers of subhalos but it is relatively small relative to the total angular momentum of the stream. The Energy-angular momentum plot of the right hand panels shows both an increasing spread of energy with increasing subhalos (top to bottom) and significantly increased structure. The relatively slim bundle of orbits in the stream with no subhalos becomes broadened with more structure as the number of subhalos increase. Each realization with subhalos is different and some show yet stronger perturbations. The spread in angular momentum leads to differing orbital periods which blurs out increasingly large features with time and the energy spread blurs out features narrower than the stream width in an orbital period.

\subsection{C-19 Model Velocity Dispersion}

\begin{figure}
\begin{center}
\includegraphics[angle=0,scale=0.55,trim=20 30 0 25, clip=true]{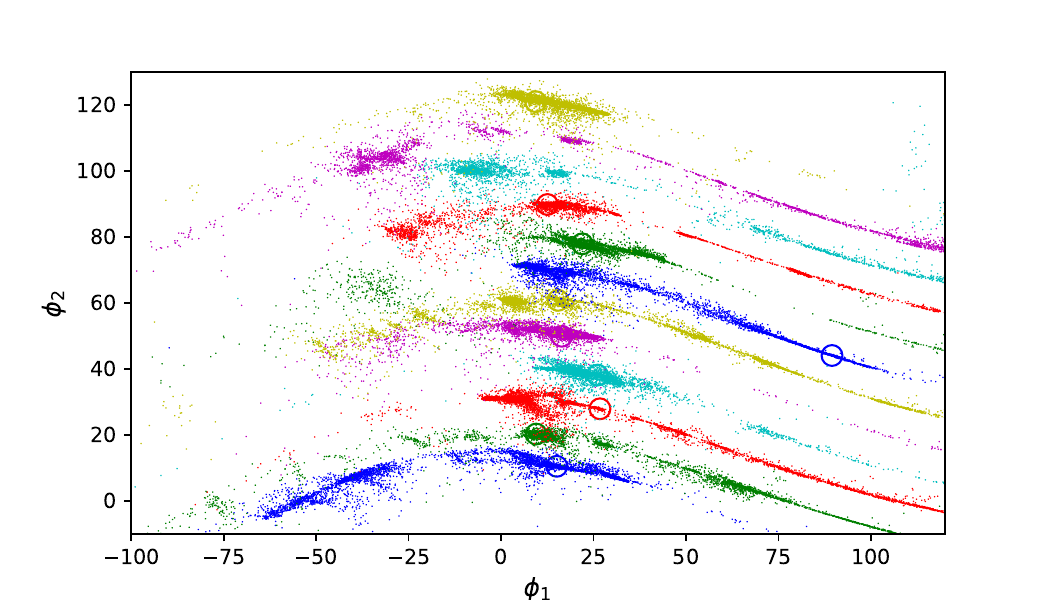}
\includegraphics[angle=0,scale=0.55,trim=20 0 0 25, clip=true]{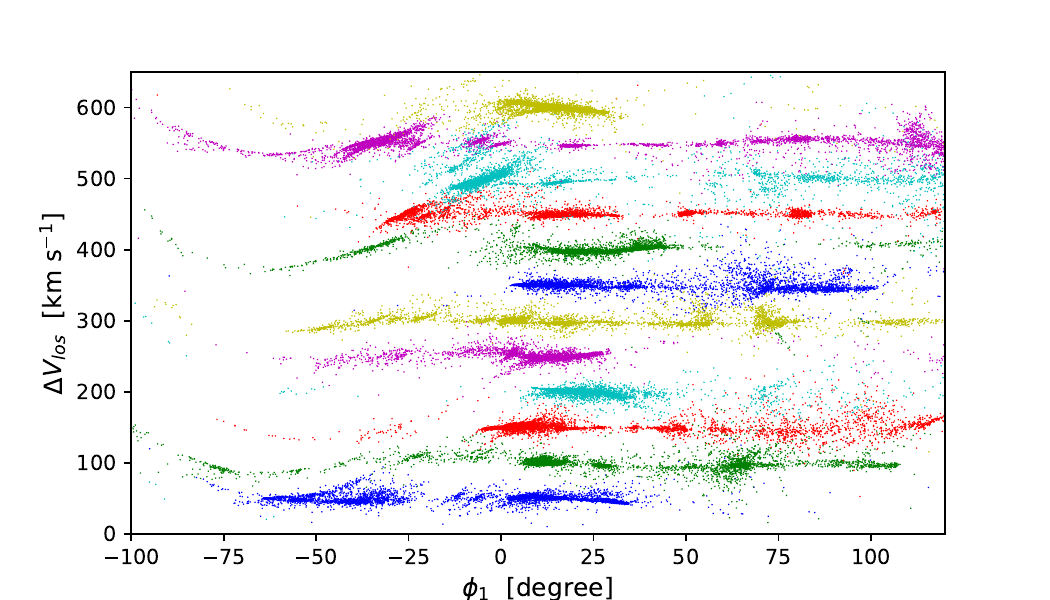}
\end{center}
\caption{The $\phi_1-\phi_2$ (top) and $\phi_1 - \Delta v_{los}$ (bottom) distribution for CDM models of C-19 with 12 rotations of 30\degree\ of the subhalo distribution. The same stream has the same color in the two panels. The streams are offset vertically 10\degree\ and 50 \kms. An approximate mean $v_{los}(\phi)$ has been subtracted from the velocities. The final location of the dissolved progenitor is marked with a large dot. The orbital apocenter is near $\phi_1(0)=$-20\degree\ and pericenter near 110\degree. }
\label{fig_svphi}
\end{figure}

The stream width and velocity dispersion vary systematically along the model streams as shown in Figure~\ref{fig_svphi}. The figure shows 12 simulations in which the subhalos are rotated 30\degree\ between simulations but are otherwise identical.  The stream coordinate $\phi_1-\phi_2$ plots (top panel) shows the stream widening near apocenter, $\phi_1=$-20\degree\ and thinning near pericenter at 110\degree. The stream widening is largely a consequence of the angular momentum spread in the orbits which have nearly the same pericenter which causes a spread of apocenter radius and orbital periods.  Although the streams thin towards pericenter the viewing angle means that the often large spread in the pericenter tangential velocities dominates the spread of line of sight velocities. Subhalos in the mass decade around $10^{7.5} M_\odot$ induce the dominant stream perturbations, of order 1 \kms. There are only a few dozen or so of these subhalos in the inner 30 kpc so there are only a few stream crossings per Gyr, leading to significant stream to stream realization differences.  Because the n-body calculation of the n-body cluster is different than the leapfrog used to calculate the progenitor orbit in the same potential, the final time dissolved progenitors are generally located past apocenter. Their locations are marked with a large dot in the top panel of Figure~\ref{fig_svphi}, where the final progenitors are spread over $\phi_1 = 25\pm 29\degree$ with the single particle orbit position of $\phi_1= -40\degree$.

\begin{figure}
\begin{center}
\includegraphics[angle=0,scale=0.45,trim=20 0 50 30, clip=true]{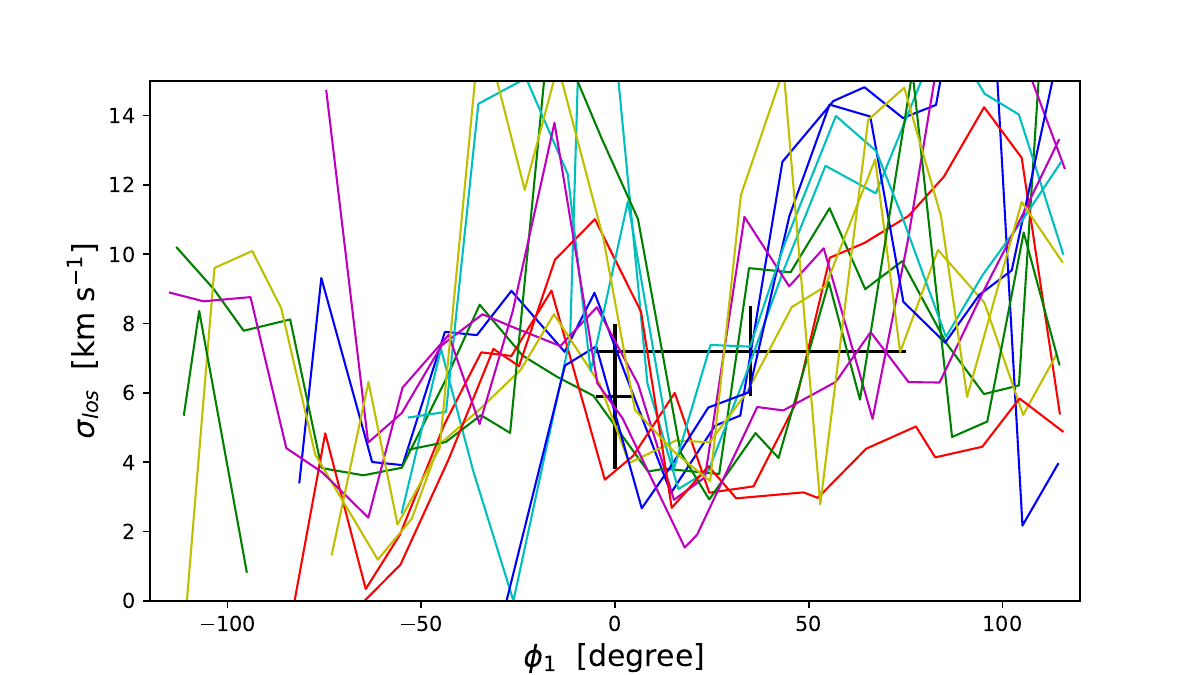}
\end{center}
\caption{The line of sight velocity dispersion for the CDM simulations started at a time of 3 Gyr at $\phi_1$ = -40\degree\ (top) with an initial half mass radius of 3.3 pc.}
\label{fig_vlos_cdm}
\end{figure}

Figure~\ref{fig_vlos_cdm} shows the line of sight velocity spread for a $2 \times 10^4 M_\sun$, $r_h=$ 3.3 pc, progenitor cluster in a simulation with CDM subhalos.  The progenitor integrated backward from $\phi_1=$-40\degree\  to a cosmological age of 3 Gyr, then the n-body cluster placed at that location and evolved forward.  The encounters between subhalos and stream particles are tracked in the calculation. Simply quadrature summing the impact approximation velocity perturbations from the individual encounters gives a velocity dispersion near 10 \kms\ for the CDM simulations. 

Table~\ref{tab_c19sim} summarizes the velocity dispersion measured in the $\phi_1=$ [-5,5] for C-19 simulations with varying progenitor masses, $M_c(0)$, initial half mass radius $r_h(0)$, subhalo populations and starting angles, $\phi_1(0)$. Larger clusters, in particular the $5\times 10^4 M_\odot$ one, have a longer half mass radius two-body relaxation time, which leads to slower mass loss. The cool stars added to the stream at late times have less heating which leads to a lower velocity dispersion.  The model with a high mass cutoff of the subhalo mass spectrum, $15\times 10^8 M_\odot$, has a velocity dispersion of $10.9 \pm 6.9 \kms$ as compared to $9.2 \pm 4.7 \kms$ for the standard model. One of the most important features is the large run-to-run simulations where the only difference is that the (starless) dark matter subhalos have been rotated 30 \degree\ from one run to the next.  Overall the simulations show that to heat C-19 to $\simeq 7 \kms$ the progenitor star cluster must orbit for $\simeq 11$ Gyr, that the orbit be in halo(s) with a CDM population of dark matter subhalos, and that the orbital phase be such that the early time orbit is effectively in the inner region where the subhalo density is high.

\begin{figure}
\begin{center}
\includegraphics[angle=0,scale=0.4,trim=50 0 0 15, clip=true]{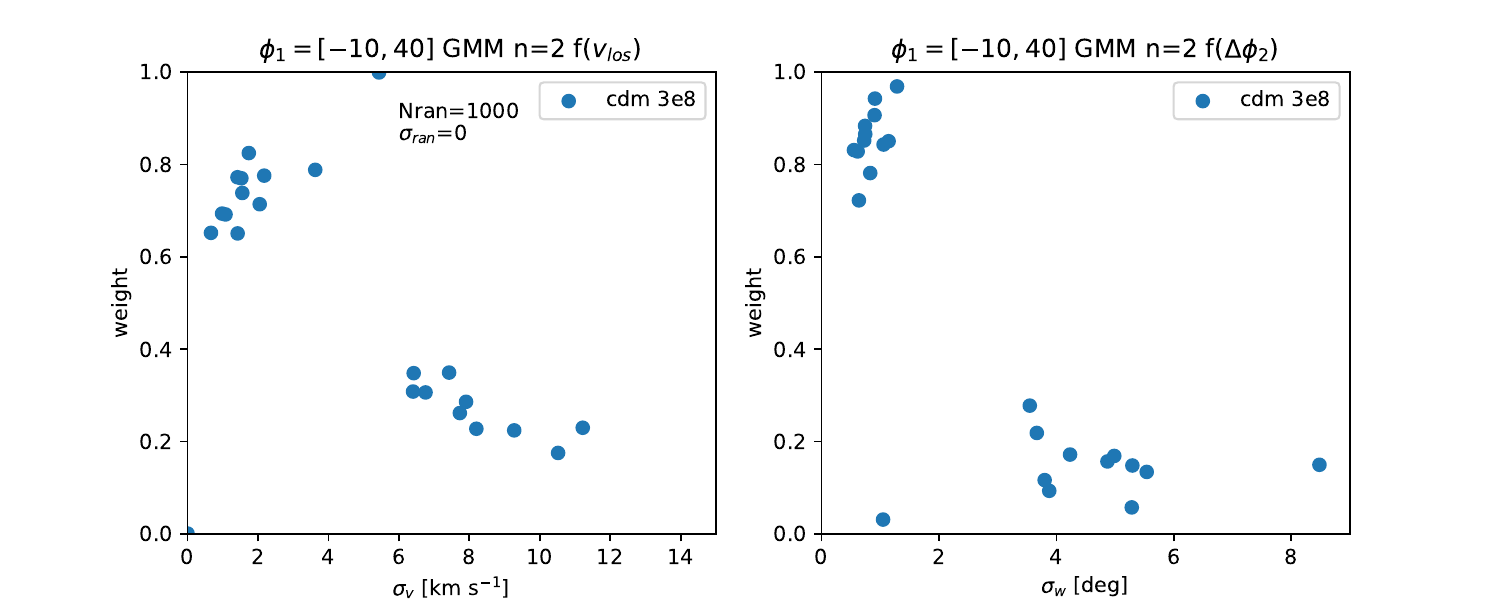}
\end{center}
\caption{Two component Gaussian mixture models  for the CDM model of the C-19 stream over $\phi_1=$ [-10,40] for CDM subhalos with an upper mass limit of $3\times 10^8$. The simulation velocity distributions are on the left and the stream width distributions on the right.}
\label{fig_gmm}
\end{figure}

Gaussian mixture models are often used to describe the core-cocoon structure of streams, as for GD-1 \citep{Valluri25} although not useful to measure for the current sparse data of C-19.  Figure~\ref{fig_gmm} shows two component Gaussian mixture models for the stream velocities and the stream width distributions for $\phi_1=$ [-10, 40] to confirm that that both thin and wide components are present in these C-19 simulations. 

\begin{table} 
\caption{Simulation Velocity Dispersions $\phi_1=$[-5, 5]
\label{tab_c19sim}}
\begin{center}
 \begin{tabular}{|r|r|r|r|r|} 
 \hline
Subhalos & $M_c(0)$ & $r_h(0)$ & $\phi_1(0)$ &$\sigma_{los}$ ~~\\
\hline
~ & $10^3 M_\odot$ & pc & deg & \kms \\
 \hline
 \multicolumn{4}{|c|}{C-19} &$5.9 \pm 2.1$ \\
\hline
none & 20 & 7.8 & -40 & 3.2 \\

$\leq 3\times 10^8$ &20 & 3.3 & -40  & $9.2\pm 4.7$ \\
$\leq 3\times 10^8$ &20 & 7.8 & -40 & $4.1 \pm 3.1$ \\
$\leq 3\times 10^8$ &10 & 3.3  & -40 & $9.3\pm 6.3$ \\
$\leq 3\times 10^8$ &20 &   3.3 &  0 &  $2.9\pm 0.8$ \\
$\leq 3\times 10^8$ &50 & 7.8 & -40 & $3.5 \pm 1.4$ \\
$\leq 15\times 10^8$ & 20 & 3.3 & -40 & $10.9 \pm 6.9$ \\
\hline
\end{tabular}
\end{center}
\end{table}

The model streams for C-19 have considerable structure along their length but the structure away from the progenitor in these model streams is less than the stream drawn from the full cosmological simulation shown in Figure~\ref{fig_simz}. Movies of two example streams  \href{https://www.astro.utoronto.ca/~carlberg/streams/417}{CDM Low Mass Clusters} show that large scale merging disperses the ``ends'' of the streams  as the main halo settles down around 7 Gyr. The primary halo in the cosmological simulation is not subject to major mergers for the last 7 Gyr of the simulation so at late times is well represented with the halo potential model used here. At the C-19 start time of 3 Gyr the dominant halo of the full cosmological simulation is about 50\% less massive, which mainly affects the orbit, not the subhalo density. The C-19 model potential does not capture the intermediate scale potential fluctuations due to accretion and merger buildup between 3 and 7 Gyr but does include the subhalo effects.

\section{Discussion and Conclusions\label{sec_discussion}}

\citet{Errani22} established, and we confirm, that the dissolution of a globular cluster in a smooth galactic halo cannot explain the width and line of sight velocity dispersion of the C-19 stream. We have shown that an old, dissolved, globular cluster stream near apocenter in a galactic halo containing CDM subhalos can explain the kinematics of the C-19 stream, under certain conditions.  Detailed modeling of C-19 using a evolving model potential and evolving subhalos drawn from a cosmological simulation of the Milky Way finds that heating C-19 requires the numbers of subhalos found in a CDM cosmology acting on the stream for $\simeq$11 Gyr. The progenitor star cluster must be sufficiently dense and low mass that it completely dissolves in $\simeq$5 Gyr. The heating process has a maximum near cosmological age of  5-7 Gyr, or a stream age of 2-4 Gyr for the starting time here. At early times,  the stream is too short to have many subhalo encounters and at late times the subhalo numbers decline so that heating is reduced. 

The orbit of the C-19 progenitor at early times needs to have a pericenter that puts it within about 10 kpc of the early Milky Way dark halo, or one of its progenitor systems, to have enough heating. Orbits that lead to early orbital pericenters of 20 kpc or more do not heat the C-19 stream to the observed value.  In as much as globular clusters form from gas in a disk-like distribution in the small galaxies present at early times, the early time orbital requirement is not unreasonable. Sufficient heating requires a CDM-like subhalo population and an orbit in the early Milky Way, or one of its to-be-accreted surrounding substructures, of 10 kpc or so. The currently limited data for C-19 and the complications of its early orbit behavior do not allow definitive conclusions about the subhalos of the Milky Way from this stream at this time. Furthermore the model for the evolving dark  subhalo population used here should be considered a first approximation to the evolution of the subhalo population in a cosmological model and could be further refined.

\begin{figure}
\begin{center}
\includegraphics[angle=0,scale=0.72,trim=0 0 0 0, clip=true]{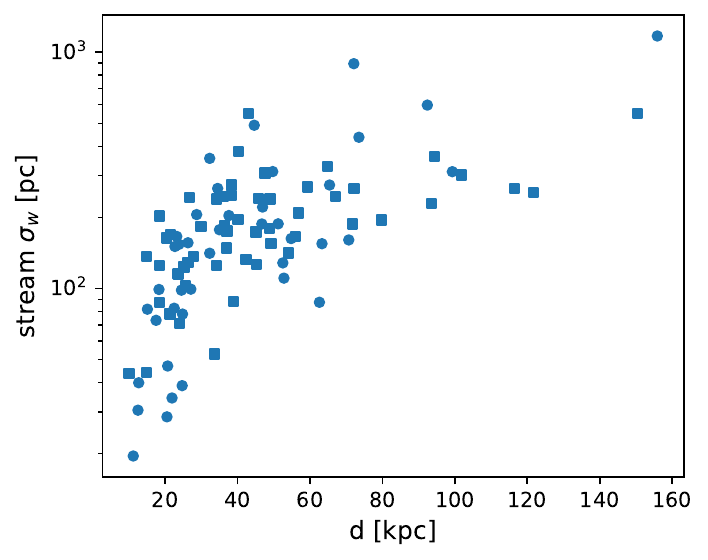}
\end{center}
\caption{CDM simulation stream widths with galoctocentric distance. Dissolved clusters are circles, remnant clusters squares. Most of the known globular cluster streams are within 30 kpc.}
\label{fig_wd}
\end{figure}

The width of C-19 is not unique among globular cluster streams. Streams  wider than 0.2 kpc  containing globular clusters  are listed in Table~\ref{tab_gc}  drawing from the \citet{Mateu23} compendium (see also \citet{BPW24}). There are a total of 27 streams wider than 0.2 kpc,  with 5 containing globular clusters and the C-19 stream. The widths are from \citet{Ibata21} and therefore use a uniform measurement approach. The wide stream of the distant cluster NGC5466 is excluded because it is likely associated with a merger remnant \citep{Malhan22} which makes the apocenter radius ambiguous.  NGC288 is a complex wide stream \citep{Grillmair24}. The Table shows that all the listed globular clusters are near the apocenters of their orbits \citep{Baumgardt19}.  On the other hand, these clusters orbit within a few kpc of the bulge, whereas C-19 is less eccentric and has a $\simeq$10 kpc pericenter. Most of the 22 other wide streams in the \citet{Mateu23} catalog have the abundance spread of a dissolved dwarf galaxy.  There should of course be other wide streams  from dissolved globular clusters at orbital apocenter but their distance and spread in position and velocity make them harder to find. The distribution of stream widths with current distance in the cosmological simulations \citep{Carlberg24} is shown in Figure~\ref{fig_wd}. About half of the simulated streams are within 30 kpc, whereas about 90\% of the Galstreams compendium \citep{Mateu23} is within 30 kpc. The radial distribution of streams will have some dependence on the radii at which globular clusters form within pre-galactic subhalos, although that dependence is  weak \citep{CK22}. 

\begin{table} 
\caption{Wide Globular Cluster Streams
\label{tab_gc}}
 \begin{center}
 \begin{tabular}{|l|r|r|r|r|r|} 
 \hline
 Name & $\quad \sigma_w $ & $\quad r_{GC}$ & $\quad r_{apo}$ & ecc\\
\hline
  &  kpc &  kpc & kpc & \\
 \hline
OmegaCen-I21         & 0.222  & 6.5 & 7.0& 0.68\\
M92-I21              & 0.231   & 9.9 & 10.5 & 0.83\\
C-19-I21             & 0.239   & 22.9 & 23.7& 0.45 \\
NGC288-I21           & 0.249  & 12.9  & 13.0& 0.59 \\
NGC1851-I21          & 0.546  &  16.7 & 19.1& 0.92\\
NGC1261-I21          & 1.042  &  18.3 & 19.3 & 0.87\\
\hline
\end{tabular}
 \end{center}
\end{table}

The hot, wide streams at apocenter become thin, cool streams at pericenter as a result of stream orbital dynamics. Stars are unbound from their progenitor cluster at pericenter in a narrow range of radii, joining the stream with a spread in velocities. The subhalo perturbations to the stream velocities are more likely to occur near orbital pericenter where the subhalo density is highest. The velocity differences lead to differences in pericenter angle, orbital tilt, and  angular momentum which causes the stream particles to spread apart near apocenter. 

The hot, wide C-19 and thin, cool, GD-1 \citep{GD1} streams cover similar radial ranges in their orbits, with the difference in their widths being largely explained as C-19 being near apocenter and GD-1 near pericenter.  Both the C-19 stream model here and the GD-1 model \citep{Carlberg25}  used the same evolving potential model, requiring that the progenitor star clusters orbit in the subhalos of the galaxy for 11 Gyr.  Both streams have some uncertainty in the progenitor mass, size and orbital phase which leads to some uncertainty in the dark matter properties. However, as more stream and better data are available the single common evolving potential model will become more tightly constrained.

\begin{acknowledgements}

KM was supported by a research grant (VIL53081) from VILLUM FONDEN. ES acknowledges funding through VIDI grant "Pushing Galactic Archaeology to its limits" (with project number VI.Vidi.193.093) which is funded by the Dutch Research Council (NWO). This research has been partially funded from a Spinoza award by NWO (SPI 78-411). This research was supported by the International Space Science Institute (ISSI) in Bern, through ISSI International Team project 540 (The Early Milky Way). 
Adrian Jenkins, Carlos Frenk and Andrew Cooper provided invaluable advice and support for computing.  This work used the DiRAC@Durham facility managed by the Institute for Computational Cosmology on behalf of the STFC DiRAC HPC Facility (www.dirac.ac.uk). The equipment was funded by BEIS capital funding via STFC capital grants ST/K00042X/1, ST/P002293/1, ST/R002371/1 and ST/S002502/1, Durham University and STFC operations grant ST/R000832/1. DiRAC is part of the National e-Infrastructure. This work used high-performance computing facilities operated by the Center for Informatics and Computation in Astronomy (CICA) at National Tsing Hua University. This equipment was funded by the Ministry of Education of Taiwan, the National Science and Technology Council of Taiwan, and National Tsing Hua University. Computations were performed on the niagara supercomputer at the SciNet HPC Consortium. SciNet is funded by: the Canada Foundation for Innovation; the Government of Ontario; Ontario Research Fund - Research Excellence; and the University of Toronto. CSF acknowledges support by the European Research Council (ERC) through Advanced Investigator grant, DMIDAS (GA 786910). ARJ and CSF acknowledge support from STFC Consolidated Grant ST/X001075/1. APC acknowledges the support of the Taiwan Ministry of Education Yushan Fellowship and Taiwan National Science and Technology Council grant 112-2112-M-007-017-MY3.
\end{acknowledgements}

\software{Gadget4: \citet{Gadget4}, Amiga Halo Finder: \citep{AHF1,AHF2}, ROCKSTAR: \citep{ROCKSTAR}, NumPy: \citep{numpy}.}

Data Availability: Final snapshots, movies, images, and example scripts are at
\href{https://www.astro.utoronto.ca/~carlberg/streams/417}{CDM Low Mass Clusters} and \href{https://www.astro.utoronto.ca/~carlberg/streams/c19}{C-19 movies}.

\bibliography{C19}{}
\bibliographystyle{aasjournal}

\end{document}